\documentclass{pnastwo}

\usepackage{cite}

\url{www.pnas.org/cgi/doi/10.1073/pnas.0709640104}
\copyrightyear{2008}
\issuedate{Issue Date}
\volume{Volume}
\issuenumber{Issue Number}

\begin{document}

\title{From Boole to Leggett-Garg: Epistemology of Bell-type Inequalities}

\author{Karl Hess\affil{1}{Center for Advanced Study, University of Illinois, Urbana, Illinois},
Hans De Raedt\affil{2}{
Zernike Institute for Advanced Materials,
University of Groningen, Nijenborgh 4, NL-9747 AG Groningen, The Netherlands
},
Kristel Michielsen\affil{3}{
Institute for Advanced Simulation, J\"ulich Supercomputing Centre,
Forschungszentrum J\"ulich, D-52425 J\"ulich,
RWTH Aachen University, D-52056 Aachen,
Germany
}
}

\contributor{{\color{red}Member submission to the Proceedings of the National Academy of Sciences
of the United States of America}}


\maketitle

\begin{article}
\begin{abstract}
In 1862, George Boole derived an inequality for variables, now known as Boolean variables, that in his opinion represents a
demarcation line between possible and impossible experience. This inequality forms an important milestone in the epistemology
of probability theory and probability measures.

In 1985 Leggett and Garg derived a physics related inequality, mathematically identical to Boole's, that according to them
represents a demarcation between macroscopic realism and quantum mechanics. Their formalism, constructed for the magnetic flux
of SQUIDS, includes general features and applies also to many other quantum experiments.

We show that a wide gulf, a wide divide,
separates the ``sense impressions" and corresponding data, as well as the postulates of macroscopic
realism, from the mathematical abstractions that are used to derive the inequality of Leggett-Garg. If the gulf can be bridged,
one may indeed derive the said inequality, which is then clearly a demarcation between possible and impossible experience: it cannot be violated and is not violated by quantum theory. We deduce from this fact that a violation
of the Leggett-Garg inequality does not mean that the SQUID-flux is not there when nobody looks, as Leggett-Garg suggest, but
instead that the probability measures may not be what Leggett-Garg have assumed them to be, when  no data can be
secured that directly relate to them. We show that similar considerations apply to other well known quantum interpretation-puzzles including
that of the two-slit experiment.
\end{abstract}

\keywords{Bell Inequality | Leggett-Garg Inequality | Foundations of Quantum Mechanics | Foundations of probability}

\dropcap{I}n 1985, Leggett and Garg~\cite{LEGG85} wrote: ``Despite sixty years of schooling in quantum mechanics, most physicists
have a very non-quantum mechanical notion of reality at the macroscopic level, which implicitly makes two assumptions. (A1)
Macroscopic realism: A macroscopic system with two or more macroscopically distinct states available to it will at all times
{\it be} in one or the other of these states. (A2) Noninvasive measurability at the macroscopic level: It is possible, in
principle, to determine the state of the system with arbitrary small perturbation on its subsequent dynamics." Leggett and Garg
continue to state that: `` ...the experimental predictions of the conjunction of (A1) and (A2) are incompatible with those of
quantum mechanics...".  We note here that Leggett and Garg later added conditions other than (A1) and (A2) which are related to
counterfactual realism. We discuss counterfactual reasoning and Bell's theorem in a separate paper~\cite{HESS15b}.

 Now, thirty years later, after ninety years of schooling in quantum mechanics, a significant body of work has been dedicated to
quantum superposition states and entanglement. Nevertheless, there are still many physicists that do not feel at ease with the
notion of quantum superposition at the macroscopic level. It is the purpose of the present paper to show that (A1) and the data
interpreted by using (A1) are still separated by a wide gulf from the mathematical abstractions and axioms used in the
derivation of the Leggett-Garg inequality. This fact was already recognized by Boole in 1862 \cite{BO1862}, who investigated the
connection between data and mathematical abstractions representing them. He showed that data could be understood by using
mathematical abstractions within a logical framework involving probability. His framework is based on using ``ultimate
alternatives" of possible outcomes, which are known since as Boolean variables. Epistemologically speaking, Boole thus linked
the data (sense impressions) with the world of ``ideas" to form a consistent whole. A necessary mathematical addition in this
``work-over" of the data are probability measures attached to the ultimate alternatives, which are initially unknowns and, as we
will see in some cases unknowable. For the latter reason, Boole asked himself the question how one could be sure, how one could
know, that one had indeed arrived at the correct ultimate possible alternatives and their probability measures.  In the course
of answering the question, he obtained his inequality. A violation of the inequality suggested to him that he had to look for
different ``ultimate alternatives" and different probability measures, in order to do justice to the complexity of the data.

 \section{Probability theory and Boole's inequality}

Boole distinguished clearly (i) events, occurrences of physical nature that are recorded as data by notebook-entries such as
$D_n$ with $n = 1, 2, 3...$ and (ii) mathematical abstractions that describe these data in the form of two-valued variables $Q_m=\pm1$
(with $m = 1, 2,3...$) representing ultimate possible alternatives, for example ``true" or ``false", up or down, head or tail and
the like. He then established a one to one correspondence of data $D_n$ and mathematical abstractions $Q_m$ and assigned yet
unknown probability measures, real numbers of the interval [0, 1], to these ultimate possible alternatives and combinations
thereof.

Boole's general procedure to eliminate these unknown probability measures to obtain the conditions of possible experience for
the $Q$s is rather complex, but he presented also an example for only three variables $Q_1, Q_2, Q_3$.\footnote{
Boole~\cite{BO1862} used x, y, z instead of $Q_1, Q_2, Q_3$ and gave on page 230 a specific example that included
the derivation of inequalities and equalities related to x, y, z.}

  In this example, he assumed the existence of probability measures $\mu, \nu...$ with $0 \leq \mu, \nu... \leq 1$ for singles,
  pairs and triples of $Q_1, Q_2, Q_3$. He then eliminated these unknowns $ \mu, \nu...$ by use of logics and algebra and obtained
  among other results the inequality:
  \begin{equation} \langle Q_1 Q_2 \rangle + \langle Q_1Q_3 \rangle + \langle Q_2Q_3 \rangle \geq -1,
  \label{26may15n2}
  \end{equation}
  where $\langle \cdot \rangle$ denotes the average. Below we also will use the notation $K_{ij} = \langle Q_iQ_j\rangle$, with $i, j = 1,
  2, 3...$ introduced by Leggett and Garg.

 This inequality for the averages, may also be re-stated for the Boolean variables themselves, if and only if the existence of
  single, as well as pair and triple joint probabilities for $Q_1, Q_2, Q_3$ is guaranteed:
  \begin{equation} Q_1 Q_2 + Q_1Q_3 + Q_2Q_3 \geq -1.
  \label{26may15n1}
  \end{equation}
  An extensive and mathematically precise derivation of the inequalities as
  related to Boole's original paper has been given in \cite{RAED11a}. Inequality Eq.~(\ref{26may15n2}) was also
  given by Leggett and Garg~\cite{LEGG85}. They assumed that the existence of all probabilities
  including single-,
 as well as joint pair- and joint triple- probabilities
  follows immediately from (A1). (A1) tells us that the results for the SQUID flux resembles coins inasmuch as these can only
  fall on heads or tails. The existence of probability measures and the use of algebra for the outcomes of experiments, however, requires much more as we are showing in detail.

In recent work~\cite{KNEE16}, it is claimed that it is already
Eq.~(\ref{26may15n1}), which is often named after Leggett-Garg, that forms according to their 1985 paper a ``constraint" to
measurement results. They do, unfortunately, not specify how the mathematical symbols $Q_1, Q_2, Q_3$ of their equation are
defined. Our symbols in Eq.~(\ref{26may15n1}) are Boole-type variables. Knee et al treat $Q_1, Q_2, Q_3$ of
Eq.~(\ref{26may15n1}) as if they were outcomes of measurements stating that Legett-Garg ``considered $Q_1, Q_2, Q_3$ as the
value taken by a macroscopic observable $Q$ measured at three consecutive times $t_1, t_2, t_3$ respectively."
Clearly, if we assume that we measure triples, such as $ Q_1=+1, Q_2=-1, Q_3=-1$,
and if we form inequalities Eq.~(\ref{26may15n1}) from these triples
(i.e. we use each measured value twice), then Eq.~(\ref{26may15n1}) follows immediately from (A1) and cannot be violated,
except if elementary algebra is violated (see also Ref.~\cite{ROSI16}).
If, on the other hand, the $Q_i$ in Eq.~(\ref{26may15n1}) are taken from different runs of experiments
but still represent values of measurements, then those with the same index need to have the same value.
Again, a violation is not possible.
Only if we use the same symbol $Q_i$ for different values (which is algebraically not permitted) may we have a violation.

From this fact we conclude that Leggett-Garg as well as Knee et al. must have considered $Q_1, Q_2, Q_3$ to be just possible
two valued outcomes of measurements, in other words some form of variables as opposed to values of these variables. In this case
the single- as well as joint pair- and triple- probabilities of $Q_1, Q_2, Q_3$ do not follow from (A1) and the claim of
Leggett-Garg that they do is false. To demonstrate this fact on the basis of Boole's work is the main topic of this paper.

Boole realized that the ``looking" at raw data and developing a view involving probability theory, involves the working-over of
the raw data and the introduction of concepts. He realized that the connection of the events of nature to numbers is nontrivial.
One must be able to treat the $Q$s in a logical fashion and subject them to the logical connections AND, OR and NOT. The $Q$s
need to be Boolean-type variables. Only in this complex way could Boole bridge the gulf between raw data and the algebra
of numbers. This bridge led also to the inequalities Eqs.~(\ref{26may15n2}) and (\ref{26may15n1}). In case of a violation of
the inequality, Boole suggested that one must question the choice of the ultimate alternatives, the existence and value of their
probability measures, or both.

In contrast to Boole, Leggett-Garg took it for granted that their $Q$s could be subjected to the logical connections AND, OR and NOT.
They also claimed that the existence of all probability measures was a direct consequence of postulate (A1).
They were then left with only one option to explain a violation of the inequality.
They abandoned macroscopic realism and questioned the nature of the $Q$s as ultimate alternatives
by involving the concept of ``superposition of states".
This concept, naturally, directed them directly to the Hilbert space of quantum theory as the only alternative, because
no algebra of numbers admits a superposition of $\pm 1$.

However, as we will see there are also ways within the framework
of macroscopic realism and ``classical" probability theory to explain or avoid inequality-violations.
These include the use of a countable infinite number of variables or indexes,
which then also necessitate the use of different probability measures.
Importantly, depending on whether pairs or triples are measured,
{\bf in the pair case}
the existence of joint triple probabilities is not necessarily guaranteed and becomes a separate question.

It took a century until Boole's inequality from 1862 was rediscovered and reformulated in a very general form by Vorob'ev~\cite{VORO62}
in 1962, who based all of his considerations on Kolmogorov's framework. Boole's variables became now Kolmogorov's
random variables with a possible range of $-1 \leq Q_1, Q_2, Q_3 \leq +1$.  Vorob'ev discovered the importance of what he called a
cyclicity when describing criteria that determine correlations of events by a Kolmogorov probability space. For the exact
meaning of the word cyclicity we refer the reader to the original work of Vorob'ev. For the reasoning in the present paper it is
sufficient to recognize the cyclicity from the following fact exhibited by Eq.~(\ref{26may15n1}): The values of the products of
variables in the first two terms determine entirely the values of these variables in the last term. This specific form of
cyclicity was already introduced in the work of Boole.

The interesting corollary of Vorob'ev's work is that given any number of variables $Q$ one always can invoke some
topological-combinatorial cyclicity that then restricts the possible choices of the correlations of a general set of data. If
one wishes to avoid such restrictions entirely, one needs to use an at least countable infinite number of variables $Q$. This is
accomplished in the Kolmogorov framework by the introduction of stochastic processes, sequences of random variables labelled by
time or space-time. The use of general space-time dependent stochastic processes to remove Vorob'ev cyclicities has been
discussed in~\cite{HESS15} and, in connection with counterfactual definiteness in \cite{HESS15b}. Below we discuss the specific
stochastic process introduced by Leggett-Garg in relation to their inequality.

\section{Connecting Boole-Leggett-Garg}

John Stuart Bell was next to discover independently an inequality similar to Eq.~(\ref{26may15n2}).
Bell's set of assumptions and postulates were different from those discussed here and
included the postulate of counterfactual definiteness~\cite{BELL93}. Counterfactual
realism was also invoked by Leggett-Garg in a defense of their original paper~\cite{LEGG10}. They used the word ``induction"
including a counterfactual explanation. We show in a separate paper~\cite{HESS15b} that the mathematically precise use of
counterfactual definiteness moves the demarcation line that the inequality represents far away from anything related to
macroscopic realism. Here, we continue to restrict ourselves to the demarcations of the inequality based on postulates (A1),
(A2).

To illustrate their quantum view of reality, and to contrast this view to the world-view of Einstein, Leggett and Garg
considered experiments involving rf-SQUIDS, superconducting quantum interference devices. These are superconducting rings
containing one or more tunneling junctions. Leggett and Garg state that quantum mechanics predicts for the magnetic flux of such
a device to oscillate back and forth between two or more macroscopically distinct states. A simpler example of a macroscopic
two-state system would be exploded and unexploded gun powder, where it is clear that the system will be most of the time in one
of the two states. From footnote 2 of Leggett and Garg~\cite{LEGG85}, one deduces that some might reason against such a simple
analogy, and there may be reasons that the detailed physics of the magnetic flux characterizing these macroscopic states of
SQUIDS is more complicated. We will not deal with these problems, however, and stipulate that Leggett and Garg indeed may
describe the magnetic flux of the SQUID by two possible macroscopic states such as the head or the tail of a coin, which
validates postulate (A1).
Claiming that the existence of all the $Q$s probability measures (including joint triple probabilities)
follows immediately from (A1), Leggett and Garg proceeded to
deduce the inequality, formally identical to Boole's as given by Eq.~(\ref{26may15n2}).

In their proof, Leggett and Garg proposed the following Gedanken-experiment that they claimed could actually be performed.
They consider many rf SQUDS, a whole ensemble of them, with unspecified location (space coordinate).  With each single one they
propose to perform a ``preparation" at time $t_0$ and subsequently measure at successive times $t_1, t_2, ....,t_i,...$ to obtain
results $Q_{t_1}=D_1, Q_{t_2}=D_2, \ldots$
at the respective times. This definition could indeed represent a general (countable
infinite) stochastic process of Kolmogorov-type and remove the Vorob'ev cyclicity (the restrictions on the correlations of the
data).  In actuality, however, Leggett-Garg use only three times $t_1, t_2, t_3$ for their proofs and do not distinguish the
SQUIDS by their location (space coordinates). They assume that for all SQUIDS of their ensemble measurements
can be performed precisely at these three times.
No distinction of the SQUIDS is permitted that results from their spacial position and possible
interactions with the other SQUIDS. All SQUIDS are being treated as identical except for the distinction with respect to the
three measurement times. In Kolmogorov's language, the data arising from all the different SQUDS are described by one stochastic
process with only three times $t_1, t_2, t_3$ to label the random variables.

\subsection{Macroscopic Reality and Leggett-Garg deductions from (A1)}

Leggett and Garg claim then: ``It immediately follows from (A1) that for an ensemble of systems prepared in some way at $t_0$,
we can define (i) joint probability densities...for $Q$ to have the values $Q_{t_i}$ at times $t_i$...(ii) correlation functions
$K_{i,j}\equiv \langle Q_{t_i}Q_{t_j}\rangle$."
Their joint probability densities involve pair and also triple probability densities for $Q_{t_1}, Q_{t_2}, Q_{t_3}$.

One must pause here for a moment to grasp the extent of this claim. Leggett and Garg do not consider how far removed the
concepts of joint probabilities (a product of human thought, human logics and mathematical definitions) are from the results of
sense impressions, from the data $D_1, D_2, D_3,...$. (A1) just tells us that there exist data arising from the recording of two
distinct sense impressions such as the heads or tails of coins and that we are sure that heads or tails are the only possible
results of our sense impressions. There is absolutely no direct connection of the data to the joint probabilities of the $Q$s,
mathematical idealizations of ultimate possible alternatives or Kolmogorov's random variables and their assumed algebra and
probability measures. Only if we have accepted all the facts of definition, logic and algebra as well as a one to one
correspondence of data and the two-valued $Q$-variables, can we possibly check whether the data are commensurate with our conceptual thinking. In
addition, there is the question what actually is measured and which ultimate alternatives describe the measurements.
Are we measuring pairs or triples? As we will see instantly, this latter question adds a significant complication.

As a preview of these complications, consider the following. From the viewpoint of quantum mechanics the SQUID is assumed to be
in a superposition of two flux-states of which one is realized with a certain probability, say one half.
Because of considerations
of entanglement, however, the probability changes as soon as the entangled partner has been measured. This shows that we are dealing
quantum mechanically with conditional probability measures that now must be represented by different ultimate alternatives.
Clearly one also needs to admit alternatives more complex than $Q_{t_1}, Q_{t_2}, Q_{t_3}$
within a classical, Einstein type framework.
The way this can be done is shown below and relates to dynamics and general stochastic processes.

\subsection{Leggett-Garg deductions from (A1) plus (A2)}

Leggett and Garg were neither aware of Boole's nor of the later work by Vorob'ev. They simply claimed that the existence of all
joint probabilities follows immediately from (A1).  The postulate (A2) just allows them to take a measurement without
influencing the system and thus they believe that (A2) links their theory to the experiments. As we can see from the details of their proposed measurements, we have the following choices:

\begin{itemize}

\item[ (i)] Measure the triples of outcomes at the three times $t_1, t_2, t_3$ and for each of the SQUIDS of the ensemble,
 which are numbered by $k= 1, 2, 3...$.
  For each triple and corresponding $k$ we have a Boole inequality:
 \begin{equation} Q_{t_1}^{k}Q_{t_2}^{k} +  Q_{t_1}^{k}Q_{t_3}^{k} + Q_{t_2}^{k}Q_{t_3}^{k} \geq -1.
 \label{23june15n1}
 \end{equation}
 Here we are using the three variables to describe precisely the $k$th SQUID at three different times.
 This inequality cannot be violated in any way, not by any
 interaction or even ``spooky'' influence.
 The inequality is, therefore, also true for the average over all $k$. In this case, the
 joint probabilities for triples are well defined and the inequality follows indeed from (A1).
 It cannot be violated and is not violated by quantum theory~\cite{RAED11a,ROSI14}.

\item[ (ii)] Measure for each SQUID just a pair of outcomes, not the triple. For example we could measure for $k = 1$ the pair of
   outcomes corresponding to $t_1$ and $t_2$, for $k =2$ the pair corresponding to $t_1$ and $t_3$ and so forth. No triples are
   measured. In this case, the Boole inequality does not follow from (A1) and it does not even follow from the conjunction of
   (A1) and (A2), because now the expectation values are determined by:
   \begin{equation} Q_{t_1}^{i} Q_{t_2}^{i} + Q_{t_1}^{m}Q_{t_3}^{m} +  Q_{t_2}^{n}Q_{t_3}^{n} \geq -3.
   \label{29may15n10}
   \end{equation}

   Here we have involved three SQUIDS of the ensemble, distinguished by the location-related
   labels $i=1,4,\ldots$, $m=2,5,\ldots$, and $n=3,6,\ldots$,
   each used for measurements at two different times.
   The joint triple events for any given SQUID are not  measured and it is unknown and unknowable
   (see physical reason below) which triple probability, if any exists at all, applies to the SQUID related to a given pair.
   Therefore, the pair correlation-functions and expectation values
   may now be arbitrarily chosen from the domain $-1 \leq K_{i,j} \leq +1$, because all the factors are, at least in principle,
   different and so are, therefore, all the ultimate alternatives and corresponding probability measures.

\end{itemize}

How can one physically justify the introduction of so many more ultimate alternatives? There are two ways of justification.
First, for each single SQUID we have chosen now one of the three combinations
$Q_{t_1}^iQ_{t_2}^i, Q_{t_1}^mQ_{t_3}^m, Q_{t_2}^nQ_{t_3}^n$
and for each of these pairs we have a different combination of time-like separations from the preparation time $t_0$. This fact opens the
possibility of a different dynamics for each of the three combinations, because of their time like differences. Second, in
addition, each SQUID is space-like separated from all other SQUIDS and is, therefore in a different
environment with respect to the possible interactions of all SQUIDS. This interaction also depends, in general, on the actual choice of time-like separations of the measurements of each SQUID and is certainly different for pair and triple measurements.

Dynamics that may be underlying such types of experiments have recently been discussed by Bush {\sl et al.}~\cite{BUSH15}. We hazard no
guess whether any of the dynamics discussed by him may be identified as actual possibilities. We emphasize, however, that according to
our understanding of the present work and that of~\cite{HESS15b} no quantum non-locality is required and all is Einstein local.

Additional physical justification of violations may be possible because of certain problems
with the noninvasive measurability at the macroscopic level, as formulated in postulate (A2), which may
clearly be violated as is well known from the Copenhagen school, because of the atomistic structure of the measurement
equipment. Even if we now measure a macroscopic flux and are able to make the measurement equipment smaller and smaller to not
disturb that flux, one may run into the inverse problem: How does the macroscopic flux of all SQUDS influence the smaller and
smaller measurement equipment? In the final analysis both the SQUID and the measurement equipment do have atomistic structure and intricate dynamics~\cite{NIEU13}
and, as even Einstein admitted \cite{EINS50}, moving away from a statistical description of physical phenomena may be
impossible.

The actual experiments proposed by Leggett-Garg have never been performed or presented to the authors knowledge. A variety of related
experiments have been performed and linked to the Leggett-Garg paper. Experiments with Caesium atoms have been presented
by Robens et al. and references to other related experiments are cited by them~\cite{ROBE15}. These experiments cannot, at least
in the opinion of the present authors, prove anything about borders between quantum and classical reality, because they use the
assumption of Leggett-Garg that probability measures for singles, pairs and triples of measurements follow immediately from
(A1). As described above there exists an enormous gulf, an enormous rift between (A1) and the probability measures of the $Q$s
and one cannot even guarantee from (A1) that the $Q$s follow some algebra such as the Boolean algebra. The $Q$s that are used by
Robens et al. may not represent the ultimate possible alternatives and an underlying dynamics may require a formidable extension of the
set of ultimate alternatives.

 \section{General quantum experiments and Leggett-Garg}

The facts discussed in (i) and (ii) above, highlight the well known conundrum of some interpretations of quantum mechanics.
If we measure triples, we cannot violate the inequality [3].
On the other hand, if we measure pairs, quantum mechanics tells us that the inequality is violated under certain circumstances.
Leggett and Garg conclude from the fact that measuring pairs may yield a violation of the inequality
and from their assumption of noninvasive measurement that any
experimentally demonstrated violation denies the ``existence" of possible outcomes
that are not measured and proves quantum superposition in our space and time system
(as opposed to being just a mathematical tool involving Hilbert space).
Therefore it appears to the authors that
Leggett-Garg have answered the question stated in the title of their paper~\cite{LEGG85} (``Is the flux there when nobody looks") in the negative.

Peres stated in the same connection ``unperformed experiments have no results".

Knowing the works of Boole and Vorob'ev, however, we do not need to go as far as doubting the presence of macroscopic or even
atomistic objects depending on whether or not we ``look".
Boole tells us for the case of a violation that we have not arrived at an understanding involving the
ultimate possible alternatives. We must choose more or different variables than just $Q_{t_1}, Q_{t_2}, Q_{t_3}$ and we may need a
stochastic process involving a countable infinite number of functions to remove all possible Vorob'ev cyclicities.
In more elementary terms, we must abandon our ideas of joint triple probability measures
when we are not measuring the data in triples but only in pairs.
We thus just may need to abandon our mathematical constructs, when the data are not commensurate with them.

The above conundrum is, of course, present in many discussions of quantum-type experiments
and is generally just not expressed as clearly as in the case of the Leggett-Garg Gedanken-experiment.
Key to the understanding of the conundrum is the fact that it arises from an unphysical invocation of joint occurrences
or events and corresponding triple probability-measures. This fact has been covered in the past from various points of view
by~\cite{PENA72,FINE74,FINE82a,BRAN87,BROD93,KUPC86,KHRE07,NIEU09,KHRE09,NIEU11,ROSI14}.

In the above example it is the assumption of the existence of sequential triple events and probability measures,
while only pair events are actually measured for each SQUID and a corresponding pair sample space must be constructed.
If triple-measurements are not made and, if instead, we deal with different sequential pair measurements at different locations,
then a dynamic interaction of all SQUIDS with different interactions in different neighborhoods leads
to Eq.~(\ref{29may15n10}) and no obvious conundrum arises.

As another example for the unphysical invocation of joint occurrences or events and corresponding joint probability measures,
consider the following known facts of the two-slit experiment.
We know that one measures an interference pattern on some detector-screen
if both slits are open and that this pattern is not even remotely equal to the sum of the patterns created
with only one slit open at a time.
One quantum-theory explanation is that the particle is only there, in one particular slit, if one is looking.
If one is not, then it is in a superposition state,
meaning in essence it exists simultaneously within the confinement of both slits.

How can we resolve this situation in terms of Boole's probability?
We must assume that there exists a dynamics that we do not observe and maybe cannot observe in its entirety.
However, the following facts alone permit an interpretation in terms of Boole-type probability theory.
A photon or electron (or any quantum entity that approaches the two slits) starts ``shaking" all
the particles constituting the material defining the slits and causes a many body ``mayhem",
a dynamics of all the involved particles and gauge fields.
It does not matter for the following argument whether we describe that shakeup
by the most modern quantum mechanical methods
or by the theoretical tools and methods that Einstein used.
This shakeup is naturally dynamically different if the particle approaches only one open slit than
it is when we have two open slits.

Denote by $Q_{t_i}^{k_1}$ the screen detection at time $t_i$ and position $s_k$ for the case when only the first slit is open.
Furthermore use $Q_{t_m}^{k_2}$ for the screen detection at time $t_m$ when only the second slit is open and finally
$Q_{t_n}^{k_{12}}$ for time $t_n$ when both slits are open. There is a fundamental problem connected to any reasoning that
combines probabilities of these occurrences in one equation, because they happen under
different physical conditions.
Yet Feynman and Hibbs \cite{FEYN65b} teach that classical physics results in a ``chance" of occurrence for the two-slit case that
equals the sum of the chances of the one slit cases. Thus they maintain that:

\begin{equation}
\langle Q_{t_i}^{k_1}\rangle  + \langle Q_{t_m}^{k_2}\rangle = \langle Q_{t_n}^{k_{12}}\rangle  ,
\label{4jan16n1}
\end{equation}
where the averages $\langle \cdot \rangle $ are taken over all $i, m, n$.

However, make the reasonable assumption that the particles and fields constituting the slit(s) exhibit an Einstein-local many
body dynamics. It is obvious then that this dynamics (including, for example, surface plasmons involving the slit material) is
different when different numbers of slits are present and we must have in general:

\begin{equation}
\langle Q_{t_i}^{k_1}\rangle  + \langle Q_{t_m}^{k_2}\rangle  \neq \langle Q_{t_n}^{k_{12}}\rangle  .
\label{4jan16n2}
\end{equation}

It is then not classical probability theory that is incompatible with quantum \cite{FEYN65b}, but it is the use of incorrect
assumptions about classical probability measures (Feynman and Hibbs used the word ``chance" without precise definition) that
leads to a conundrum~\cite{BALL86}. Note that our notation is based on Boole-Kolmogorov with a corresponding space of events in
mind. The difference of having one and two slits open plays, therefore, a major role~\cite{KHRE99}.

Our reasoning just is that one cannot oversimplify or distort macroscopic realism by assuming
the existence of measurements conditional to the joint occurrence of events that cannot jointly occur
(thus technically involving conditioning on impossible events, a probabilistic no no).
To avoid contradictions and quantum nonlocalities (as opposed to Einstein locality),
one further needs to include dynamic many body
interactions in the slit material that quantum theory does not require,
because it somehow efficiently circumvents them by using probability amplitudes
and Hilbert space instead of probability measures.

If we attempt to develop a space-time picture without involving Hilbert space, we must admit general many body stochastic
processes, in order to avoid conundrums of the kind described above. We are convinced that most physicists will agree that one
may regard the slit material as a many body quantum mechanical system, consisting of particles and fields that interact with the
incoming particles and fields. Naturally, we understand that the complete (classical or quantum) many-body treatment of
equipment and environment (including the observer) leads to an infinite regress that has to be cut-off like a Gordian knot at
some point. It is, however, also our opinion that imposing such a cut-off on ``classical" type of thinking when performing
Leggett-Garg-type proofs, is leading physics down the wrong path and presents conundrums that are artificial.

The authors are convinced that these conundrums are a consequence of some hidden dynamics that is in its essence covered by the
formalism of quantum mechanics which provides us with suitable long term averages in spite of the very simplified description of
the measurement equipment. The conundrum arises only if we ``derive" from these long term averages probability measures for certain oversimplified
variables (that do not represent the ultimate logical alternatives that some believe they are). It is not that the alternatives do
not exist when we are not looking. We maintain that the probability measures that are assigned in Leggett-Garg-type proofs to
these supposedly ultimate alternatives are just incorrect and a much larger number as well as more complicated alternatives will exist and must be used.

Leggett and Garg did realize soon after publication of their 1985 paper~\cite{LEGG85} that their reasoning was not air-tight and
proposed later the inclusion of another postulate that is related to counterfactual realism \cite{LEGG10}. The problems with this
postulate are discussed in a separate paper~\cite{HESS15b}.

\section{Conclusion}

Thus, if we follow the probability theory of Boole carefully, no quantum conundrum arises. We only need to postulate that we
have not arrived at the precise ultimate alternatives and their probability measures. Different physics related labels may
characterize the $Q$s and we cannot attach probability measures and postulate the existence of joint probabilities if the data
do not support the existence of such joint probabilities. The same reasoning applies to approaches using the probability theory
of Kolmogorov. We just need to remove the Vorob'ev cyclicities by use of stochastic processes and introduce additional space and
time (or space-time) labels for the $Q$s in order to avoid conflicts with quantum theory.

For the example of the two-slit experiment, this means that we cannot assign probability measures for detection on a screen by
considering each single slit separately and independently. If both slits are open, the many body interactions of incoming
particles and fields with the quantum entities of the slit-material necessitate different indexing and time labels for the cases
of two slits simultaneously open or just one slit open during separate and different time periods. The question through which
slit a particle propagates is more appropriately replaced by the question which many body interactions the incoming particle
undergoes.

We are convinced that the well known ``one liner" of Peres ``unperformed experiments have no results" must be replaced by: ``
For unperformed experiments, we cannot assign probability measures to oversimplified ultimate alternatives (Kolmogorov random
variables), or even postulate the existence of these probability measures, which may have nothing to do with macroscopic
realism".

\begin{acknowledgments}
The authors wish to express their gratitude to Andrei Khrennikov and Theo Nieuwenhuizen for many valuable suggestions.
\end{acknowledgments}

\bibliography{/d/papers/all16}   

\end{article}

\end{document}